\documentclass[aps,prb,floatfix,twocolumn,showpacs]{revtex4}
\usepackage{amsmath}
\usepackage{graphicx}
\usepackage{amsfonts}
\usepackage{amssymb}
\begin{document}

\title{Conductance and current noise of a superconductor/ferromagnet quantum point contact}

\author{Audrey Cottet}
\affiliation{Laboratoire de Physique Th\'{e}orique et Hautes \'{E}nergies, Universit\'{e}s Paris 6 et 7, CNRS, UMR 7589, 4 place Jussieu, F-75252 Paris Cedex 05, France}

\author{Wolfgang Belzig}
\affiliation{Fachbereich Physik, Universit\"at Konstanz, D-78457 Konstanz, Germany}


\begin{abstract}
We study the conductance and current noise of a
superconductor/ferromagnet ($S/F$) single channel Quantum Point Contact
(QPC) as a function of the QPC bias voltage, using a scattering approach. We
show that the Spin-Dependence of Interfacial Phase Shifts (SDIPS) acquired
by electrons upon scattering by the QPC can strongly modify these signals.
For a weakly transparent contact, the SDIPS induces sub-gap resonances
in the conductance and differential Fano factor curves of the QPC. For
high transparencies, these resonances are smoothed, but the shape
of the signals remain extremely sensitive to the SDIPS. We show that noise
measurements could help to gain more information on the device, e.g. 
in cases where the SDIPS modifies qualitatively the
differential Fano factor of the QPC but not the conductance.
\end{abstract}

\date{\today}
\maketitle

\section{Introduction}

Mesoscopic circuits with ferromagnetic elements are generating a growing
interest, both for the fundamentally new effects they can exhibit due to the
lifting of spin-degeneracy, and for the possibilities of technological
advances, e.g. towards nanospintronics\cite{Prinz}. The description of the
interface between ferromagnetic and non-magnetic elements is crucial to
understand the behavior of these devices. Obviously, one first has to take
into account the spin-polarization of the electronic transmission
probabilities through the interfaces, a property which generates the widely
known magnetoresistance effects\cite{MR}. Importantly, the phases of the
scattering amplitudes also depend on the spin in general. This so-called
spin-dependence of interfacial phase shifts (SDIPS) is less frequently taken
into account. However, the SDIPS has already been found to affect the behavior
of many different types of mesoscopic conductors with ferromagnetic leads,
like diffusive normal metal islands \cite{FNF}, resonant systems
\cite{CottetEurophys,SST}, Coulomb blockade systems\cite{CB, Cottet06,SST}, or
Luttinger liquids\cite{Luttinger}. The SDIPS can also have numerous types of
consequences in superconducting/ferromagnetic ($S/F$) hybrid systems
\cite{Tokuyasu,otherBC,AbsSpinValve,cottet05,cottet06,Braude,Zhao}. It can
e.g. produce effective field effects in superconducting
electrodes\cite{Tokuyasu, AbsSpinValve,cottet06}, or a phase shift of the
spatial oscillations of the superconducting correlations induced by the
proximity effect in a diffusive $F$ electrode\cite{cottet06,cottet05}. The
SDIPS has also been predicted to induce triplet correlations in a $S$
electrode in contact with several ferromagnetic insulators with non-collinear
polarization directions\cite{Braude}.

Recently, mean current measurements through $S/F$ quantum points contacts
(QPCs) have raised much interest as a possible means to determine the
polarization of a $F$ material. However, the effects of the SDIPS in these
devices have not raised much attention so far. In most cases, the experimental
results were interpreted in terms of a generalization of the
Blonder-Tinkham-Klapwick (BTK) model\cite{BTK} to the spin-dependent case
\cite{Soulen}. We would like to emphasize that the description of a $S/F$
interface in terms of a delta-function barrier naturally includes a finite
SDIPS but does not allow one to analyze its effects separately. Moreover, the
SDIPS included in the generalized BTK model is specific for the delta-function
barrier, and for a more realistic interface potential the SDIPS can be
different. This is the reason why the influence of the transmission
probabilities and of the SDIPS have to be investigated separately. The
scattering approach seems perfectly adapted for this purpose since it allows
to account explicitly for scattering phases\cite{Blanter}. In a few works,
current transport through $S/F$ QPCs was described with a single-channel
scattering description\cite{PerezWillard,Martin-Rodero}. However, the SDIPS
was not taken into account in these works.

In this article, we study how the conductance and the current noise through a
single channel $S/F$ QPC are affected by the SDIPS, using a scattering
approach. We would like to emphasize that the current noise in hybrid $S/F$
systems has draw little attention so far\cite{NoiseSF}, although it can
provide more information on the system considered. We find that the behavior
of the QPC depends on its spin-dependent normal state transmission
probabilities $T_{\sigma}$, but also on the difference $\Delta\varphi$ between
the normal state reflection phases of up spins and down spins against $F$. The
device is thus more difficult to characterize than in the spin-degenerate case
since there are two more parameters to determine, i.e. $T_{\uparrow
}-T_{\downarrow}$ and $\Delta\varphi$. We show that a finite SDIPS
$\Delta\varphi\neq0$ can strongly modify the behavior of the QPC, for instance
by shifting the Andreev peaks appearing in the QPC conductance to subgap
voltages, as found by Zhao et al. for $S/F/N$ quantum point
contacts\cite{Zhao}, or by producing subgap peaks or dips in the differential
Fano factor curves. At finite temperatures, we find that the SDIPS-induced
shift of the Andreev peaks in the conductance can be difficult to distinguish
from a gap reduction. We show that, in this context, a noise measurement can
be very useful to characterize better the contact properties. This is
particularly clear in cases where the QPC differential Fano factor versus bias
voltage is qualitatively modified by the SDIPS while the conductance curve
remains similar to the curves obtained without a SDIPS.

This paper is organized as follows: section II.A introduces expressions of the
conductance and noise of a $S/F$ interface in the scattering formalism,
section II.B presents an explicit expression of the scattering matrix of a
single channel $S/F$ QPC, section II.C discusses the conductance and noise of
this QPC at zero temperature, and section II.D discusses the finite
temperature case. Section III concludes.

\section{Conductance and noise of a single channel $S/F$ QPC}

\subsection{Current and noise through a $S/F$ interface in the scattering approach}

We consider a superconducting/ferromagnetic ($S/F$) interface with $S$ a
conventional BCS conductor with a s-wave symmetry. We assume that the $S$ and
$F$ leads can be considered as ballistic, so that the $S/F$ interface can be
described with the scattering approach\cite{Blanter} and the leads with
Bogoliubov-De Gennes (BdG) equations\cite{BdG}. We use a gauge transformation
to set the origin of the quasiparticles energies to the Fermi level of $S$,
i.e. $\mu_{S}=0$\cite{Datta}. Due to the bias voltage $V$ applied to the QPC,
the Fermi level of $F$ is $\mu_{F}=-eV$, with $e=\left|  e\right|  $ the
absolute value of the electron charge. In the following, we will denote by
$(e,\sigma)$ electron states with spin $\sigma$ and by $(h,-\sigma)$ hole
states in the $-\sigma$ spin band, with $\sigma\in\{\uparrow,\downarrow\}$ a
spin component collinear to the polarization of $F$. A hole $(h,-\sigma)$
corresponding to an empty electronic state at an energy $-\varepsilon$ carries
the same energy $\varepsilon$ as an electron $(e,\sigma)$ occupying a state
with energy $\varepsilon$. These two types of quasiparticles are coupled by
Andreev reflection processes occurring at the $S/F$ interface. One can thus
write
\[
b_{M,\alpha}(\varepsilon)=\sum\nolimits_{\gamma\in\mathcal{E}_{\sigma}%
,Q\in\left\{  F,S\right\}  }\mathcal{S}_{M,Q}^{\alpha,\gamma}(\varepsilon
)a_{Q,\gamma}(\varepsilon)
\]
for $\alpha\in\mathcal{E}_{\sigma}$, with $\mathcal{E}_{\sigma}=\{(e,\sigma
),(h,-\sigma)\}$ and $\mathcal{S}(\varepsilon)$ the scattering matrix of the
$S/F$ interface for quasiparticles carrying an energy $\varepsilon$. Here, we
consider the single channel case, so that $a_{M,\alpha}(\varepsilon)$
[$b_{M,\alpha}(\varepsilon)$] refers to the annihilation operator associated
to the incident [outgoing] state for a particle with type $\alpha
\in\mathcal{E}_{\sigma}$ and energy $\varepsilon$ of lead $M$. In the
following, we use a picture consisting of both positive and negative energy
states. The operator $\hat{I}(t)$ associated to the total current flowing
through the device at time $t$ can be calculated as the sum of the electron
and hole currents of the $\mathcal{E}_{\sigma}$ space, which replaces the
summation on the spin direction\cite{Datta}. More precisely, one has $\hat
{I}(t)=\hat{I}_{e,\sigma}(t)+\hat{I}_{h,-\sigma}(t)$ with
\[
\hat{I}_{\gamma}(t)=\frac{i\hbar e}{2m}\int dr_{\perp}\left\{  \psi_{F,\gamma
}^{\dagger}\left(  \frac{\partial\psi_{F,\gamma}}{\partial z}\right)  -\left(
\frac{\partial\psi_{F,\gamma}^{\dagger}}{\partial z}\right)  \psi_{F,\gamma
}\right\}
\]
for $\gamma\in\mathcal{E}_{\sigma}$. Here, $z$ is the coordinate along the
leads and $r_{\perp}$ is the transverse coordinate. The field operator
$\psi_{F,\gamma}$ associated with particles of type $\gamma$ in lead $F$ is
defined as
\begin{align*}
\psi_{F,\gamma}  &  =\int\limits_{-\infty}^{+\infty}d\varepsilon
e^{-\frac{i\varepsilon t}{\hbar}}\frac{\chi_{F(r_{\perp})}}{\sqrt{2\pi
v_{F}(\varepsilon)}}\\
&  \times(a_{F,\gamma}(\varepsilon)e^{i\lambda\lbrack\gamma]k_{F}%
(\varepsilon)z}+b_{F,\gamma}(\varepsilon)e^{-i\lambda\lbrack\gamma
]k_{F}(\varepsilon)z})
\end{align*}
with $\lambda\lbrack(e,\sigma)]=-1$, $\lambda\lbrack(h,-\sigma)]=+1.$ We have
introduced above different quantities which characterize the conduction
channel of the QPC, i.e. the transverse wave function $\chi_{F}$, the
wavevector $k_{F}$, and the velocity of carriers $v_{F}=\hbar k_{F}/m$. The
operator conjugated to $\psi_{F,\gamma}$ is denoted $\psi_{F,\gamma}^{\dagger
}$. Neglecting the energy dependence of $v_{F}$ (see e.g. Ref. \onlinecite
{Blanter}), one finds%
\begin{align}
\hat{I}_{\gamma}(t)  &  =\frac{e}{h}\lambda\lbrack\gamma]\iint\limits_{-\infty
,-\infty}^{+\infty,+\infty}d\varepsilon_{1}d\varepsilon_{2}e^{\frac
{i(\varepsilon_{1}-\varepsilon_{2})t}{\hbar}}\nonumber\\
&  \times\sum\limits_{\substack{\alpha_{1},\alpha_{2}\\M_{1},M_{2}}%
}a_{M_{1},\alpha_{1}}^{\dagger}(\varepsilon_{1})\mathcal{A}_{F,M_{1},M_{2}%
}^{\gamma,\alpha_{1},\alpha_{2}}(\varepsilon_{1},\varepsilon_{2}%
)a_{M_{2},\alpha_{2}}(\varepsilon_{2}) \label{Iinst}%
\end{align}
with%
\begin{align}
\mathcal{A}_{F,M_{1},M_{2}}^{\gamma,\alpha_{1},\alpha_{2}}(\varepsilon
_{1},\varepsilon_{2})  &  =\mathbb{I}_{F,\gamma}\delta_{M_{1},F}\delta
_{M_{2},F}\delta_{\gamma,\alpha_{1}}\delta_{\gamma,\alpha_{2}}\nonumber\\
&  -\left(  \mathcal{S}_{F,M_{1}}^{\gamma,\alpha_{1}}(\varepsilon_{1})\right)
^{\dag}\mathcal{S}_{F,M_{2}}^{\gamma,\alpha_{2}}(\varepsilon_{2}) \label{A}%
\end{align}
In the above Eqs., capital Latin indices correspond to the lead $F$ or $S$,
Greek indices correspond to the electron or hole band of the space
$\mathcal{E}_{\sigma}$, and $\mathbb{I}_{F,\gamma}$ is the identity matrix in
the subspace of states of type $\gamma$ of lead $F$. In this paper, we study
the average current $\left\langle I\right\rangle $ flowing through the
interface and the zero-frequency current noise $S=2\int_{-\infty}^{+\infty
}dt\left\langle \left[  \hat{I}(t)-\langle I\rangle\right]  \left[  \hat
{I}(0)-\langle I\rangle\right]  \right\rangle $. Equations (\ref{Iinst}) and
(\ref{A}) lead to the expressions \cite{DeJong, Anantram}%
\begin{equation}
\left\langle I\right\rangle =\frac{e}{h}\sum\limits_{M,\alpha,\gamma}%
\lambda(\alpha)\int\nolimits_{-\infty}^{+\infty}d\varepsilon f_{M}^{\gamma
}(\varepsilon)\mathcal{A}_{F,M,M}^{\alpha,\gamma,\gamma}(\varepsilon
,\varepsilon) \label{Igal}%
\end{equation}
and
\begin{align}
S(V)  &  =\frac{2e^{2}}{h}\sum\limits_{M_{1},M_{2},\gamma_{1},\gamma
_{2},\alpha,\beta}\lambda(\alpha)\lambda(\beta)\nonumber\\
&  \times\int\nolimits_{-\infty}^{+\infty}d\varepsilon f_{M_{1}}^{\gamma_{1}%
}(\varepsilon)(1-f_{M_{2}}^{\gamma_{2}}(\varepsilon))\mathcal{A}%
_{F,M_{1},M_{2}}^{\alpha,\gamma_{1},\gamma_{2}}(\varepsilon,\varepsilon
)\mathcal{A}_{F,M_{2},M_{1}}^{\beta,\gamma_{2},\gamma_{1}}(\varepsilon
,\varepsilon) \label{Sgal}%
\end{align}
We have introduced above the Fermi factors $f_{M}^{\gamma}(\varepsilon
)=\left(  1+\exp[\left(  \varepsilon+\lambda\lbrack\gamma]\mu_{M}\right)
/k_{B}T]\right)  ^{-1}$ with $T$ the temperature. In the limit $V\rightarrow
0$, Eqs. (\ref{Igal}) and (\ref{Sgal}) fulfill the fluctuation dissipation
theorem, i.e. $S=4k_{B}TG$ with $G(V)=\partial I/\partial V$\textbf{.}

In the limit $T=0$, simplified expressions of $G$ and $S$ can be obtained. In
order to account for the two spin bands in the same way, we use a symmetry
property stemming from the structure of the BdG equations, i.e. $\mathcal{T}%
_{M_{1},M_{2}}^{\alpha,\beta}(-\varepsilon)=\mathcal{T}_{M_{1},M_{2}%
}^{\widetilde{\alpha},\widetilde{\beta}}(\varepsilon)$, with $(\alpha
,\beta)\in\mathcal{E}_{\sigma}^{2}$, $(M_{1},M_{2})\in\{S,F\}^{2}$,
$\widetilde{(e,\sigma)}=(h,\sigma)$ and $\widetilde{(h,-\sigma)}=(e,-\sigma)$
(see derivation in Appendix A). Equations (\ref{Igal}) and (\ref{Sgal})
combined with the unitarity of $\mathcal{S}(\varepsilon)$ then lead to
\begin{equation}
G\left(  V\right)  =\frac{e^{2}}{h}\sum\limits_{\sigma}\mathcal{W}%
_{F,F}^{(e,\sigma)}(\varepsilon=-eV) \label{Gfns}%
\end{equation}
with
\begin{equation}
\mathcal{W}_{F,F}^{(e,\sigma)}(\varepsilon)=\mathbb{I}_{F,(e,\sigma
)}-\mathcal{T}_{F,F}^{(e,\sigma),(e,\sigma)}(\varepsilon)+\mathcal{T}%
_{F,F}^{(h,-\sigma),(e,\sigma)}(\varepsilon) \label{W}%
\end{equation}
and\cite{comp}
\begin{align}
S(V)  &  =2\frac{e^{2}}{h}\sum\limits_{\sigma}\int\nolimits_{\mathcal{D}_{V}%
}d\varepsilon\left[  2\mathcal{T}_{F,F}^{(h,-\sigma),(e,\sigma)}%
(\varepsilon)\mathcal{T}_{F,F}^{(e,\sigma),(e,\sigma)}(\varepsilon)\right.
\nonumber\\
&  +\mathcal{T}_{F,F}^{(e,\sigma),(e,\sigma)}(\varepsilon)\left(
\mathbb{I}_{F,(e,\sigma)}-\mathcal{T}_{F,F}^{(e,\sigma),(e,\sigma
)}(\varepsilon)\right) \nonumber\\
&  \left.  +\mathcal{T}_{F,F}^{(h,-\sigma),(e,\sigma)}(\varepsilon)\left(
\mathbb{I}_{F,(h,-\sigma)}-\mathcal{T}_{F,F}^{(h,-\sigma),(e,\sigma
)}(\varepsilon)\right)  \right]  \label{Sfns}%
\end{align}
with $\mathcal{T}_{M_{1},M_{2}}^{\gamma_{1},\gamma_{2}}(\varepsilon)=\left|
\mathcal{S}_{M_{1},M_{2}}^{\gamma_{1},\gamma_{2}}(\varepsilon)\right|  ^{2}$.
Note that Eq. (\ref{Gfns}) is valid only if $\mathcal{W}_{F,F}^{\gamma
}(\varepsilon)$, with $\gamma\in\mathcal{E}_{\sigma}$, can be considered as
independent from $V$. The integration domain $\mathcal{D}_{V}$ in
Eq.(\ref{Sfns}) is $[-e\left|  V\right|  ,0]$\ for $V>0$ and $[0,e\left|
V\right|  ]$ for $V<0$, similarly to the fact that the elements $\mathcal{W}%
_{F,F}^{(e,\sigma)}$ contributing to $G\left(  V\right)  $ in Eq. (\ref{Gfns})
must be taken at $\varepsilon=\mp e\left|  V\right|  $ depending on the sign
of $V$.

Before concluding this section, we note that in the multichannel case, the
elements $\mathcal{S}_{M,Q}^{\alpha,\gamma}(\varepsilon)$ would be matrices
relating the incoming and outgoing states of the different channels of leads
$Q$ and $M$. In this case, one could generalize straightforwardly Eqs.
(\ref{Igal}), (\ref{Sgal}), (\ref{Gfns}) and (\ref{Sfns}) by applying to their
right-hand sides a trace on the channels space and using $\mathcal{T}%
_{M_{1},M_{2}}^{\gamma_{1},\gamma_{2}}(\varepsilon)=\left(  \mathcal{S}%
_{M_{1},M_{2}}^{\gamma_{1},\gamma_{2}}(\varepsilon)\right)  ^{\dag}%
\mathcal{S}_{M_{1},M_{2}}^{\gamma_{1},\gamma_{2}}(\varepsilon)$. In section
II, we focus on the single channel case. We briefly discuss possible
extensions of this work to the multichannel case in section III.

\subsection{Scattering matrix of a single channel $S/F$ QPC}

We consider a $S/F$ interface which is narrow compared to the coherence length
of $S$, so that it can be modeled as a specular $S/N$ interface in series with
a (possibly) dirty $N/F$ interface, with the length of $N$ tending to
zero\cite{BeenakkerSNN}. We would like to emphasize that the $N$ layer is
introduced artificially, it is merely a trick to describe the superconducting
interface on a scale much shorter that the coherence length. The matrix
$\mathcal{S}(\varepsilon)$ of the $S/F$ interface can expressed in terms of
the scattering matrix $\mathcal{P}^{\sigma}(\varepsilon)$ of electrons with
spins $\sigma$ on the $N/F$ interface and of the amplitude $\gamma$ of the
Andreev reflections at the $S/N$ interface. One finds in particular
\cite{Xia},
\begin{align}
\mathcal{S}_{F,F}^{(e,\sigma),(e,\sigma)}(\varepsilon)  &  =\mathcal{P}%
_{FF}^{\sigma}(\varepsilon)\nonumber\\
&  +\gamma^{2}\mathcal{P}_{FN}^{\sigma}(\varepsilon)N^{\sigma}\left(
\mathcal{P}_{NN}^{-\sigma}(-\varepsilon)\right)  ^{\ast}\mathcal{P}%
_{NF}^{\sigma}(\varepsilon) \label{S2}%
\end{align}%
\begin{equation}
\mathcal{S}_{F,F}^{(h,-\sigma),(e,\sigma)}(\varepsilon)=\gamma\left(
\mathcal{P}_{FN}^{-\sigma}(-\varepsilon)\right)  ^{\ast}M^{\sigma}%
\mathcal{P}_{NF}^{\sigma}(\varepsilon) \label{S1}%
\end{equation}%
\begin{equation}
M^{\sigma}=\left[  \mathbb{I}_{N,(e,\sigma)}-\gamma^{2}\mathcal{P}%
_{NN}^{\sigma}(\varepsilon)\left(  \mathcal{P}_{NN}^{-\sigma}(-\varepsilon
)\right)  ^{\ast}\right]  ^{-1}%
\end{equation}%
\begin{equation}
N^{\sigma}=\left[  \mathbb{I}_{N,(h,-\sigma)}-\gamma^{2}\left(  \mathcal{P}%
_{NN}^{-\sigma}(-\varepsilon)\right)  ^{\ast}\mathcal{P}_{NN}^{\sigma
}(\varepsilon)\right]  ^{-1} \label{S4}%
\end{equation}
From Eqs. (\ref{Gfns}) and (\ref{Sfns}), at $T=0$, calculating the current and
the noise through the $S/F$ interface only requires one to know the two
elements of $\mathcal{S}(\varepsilon)$ given above. For $T\neq0$, one must use
Eqs. (\ref{Igal}) and (\ref{Sgal}), so that the whole $S(\varepsilon)$ matrix
is necessary (the other elements of $S(\varepsilon)$ are given in appendix B).

There remains to introduce explicit expressions for $\gamma$ and
$\mathcal{P}^{\sigma}$. First, the amplitude $\gamma$ can be calculated using
the ballistic BdG equations to model the $S/N$ interface\cite{BTK}, with a
step approximation for the gap $\Delta$ of $S$. This gives $\gamma
=(\varepsilon-\mathrm{i}\sqrt{\Delta^{2}-\varepsilon^{2}})/\Delta$ for
$\left|  \varepsilon\right|  <\Delta$ and $\gamma=(\varepsilon-\mathrm{sgn}%
(\varepsilon)\sqrt{\varepsilon^{2}-\Delta^{2}})/\Delta$ for $\left|
\varepsilon\right|  >\Delta$. Secondly, in the single channel case, the
unitary of $\mathcal{P}^{\sigma}$ leads to
\[
\mathcal{P}^{\sigma}=\left[
\begin{tabular}
[c]{ll}%
$\sqrt{R_{\sigma}}\exp\left[  i\varphi_{FF}^{\sigma}\right]  $ &
$\sqrt{T_{\sigma}}\exp\left[  i\varphi_{FN}^{\sigma}\right]  $\\
$\sqrt{T_{\sigma}}\exp\left[  i\varphi_{NF}^{\sigma}\right]  $ &
$\sqrt{R_{\sigma}}\exp\left[  i\varphi_{NN}^{\sigma}\right]  $%
\end{tabular}
\ \ \ \ \right]
\]
with $R_{\sigma}+T_{\sigma}=1$ and $\varphi_{NN}^{\sigma}+\varphi_{FF}%
^{\sigma}=\varphi_{NF}^{\sigma}+\varphi_{FN}^{\sigma}+\pi~[2\pi]$. In
principle, the scattering phases $\varphi_{ij}^{\sigma}$, with $(i,j)\in
\{N,F\}^{2}$, are spin-dependent because electrons are affected by a
spin-dependent scattering potential at the $S/F$ interface. For simplicity, we
will assume that $R_{\sigma}$, $T_{\sigma}$ and $\varphi_{ij}^{\sigma}$ are
independent from $\varepsilon$ and $V$, which ensures the validity of Eq.
(\ref{Gfns}) (see Ref. \onlinecite{Vdrop}). Note that the expression that we
have introduced in this section for $\mathcal{S}$ implies $G(V)=G(-V)$ and
$S(V)=S(-V)$, which is not a general property of superconducting hybrid
devices (see e.g. Ref.~\onlinecite{Lesovik}).

\subsection{Conductance and noise of the QPC at zero-temperature}

This section discusses the conductance and current noise of the single channel
$S/F$ QPC at zero temperature ($T=0$). For $e\left|  V\right|  <\Delta$, no
quasiparticle propagates in $S$, so that the unitarity of $\mathcal{S}%
(\varepsilon)$ leads to $\mathcal{T}_{F,F}^{(e,\sigma),(e,\sigma
)}=1-\mathcal{T}_{F,F}^{(h,-\sigma),(e,\sigma)}$. One can thus calculate $G$
and $S$ from Eqs. (\ref{Gfns}) and (\ref{Sfns}) by using
\begin{align}
&  \mathcal{T}_{F,F}^{(h,-\sigma),(e,\sigma)}(\varepsilon)\nonumber\\
&  =\frac{T_{\uparrow}T_{\downarrow}}{1+R_{\uparrow}R_{\downarrow}%
-2\sqrt{R_{\uparrow}R_{\downarrow}}\cos[\varphi_{NN}^{\sigma}-\varphi
_{NN}^{-\sigma}+2\varphi_{a}]} \label{Thesingle}%
\end{align}
for $\left|  \varepsilon\right|  <\Delta$, with $\varphi_{a}(\varepsilon
)=\arg[\gamma]$. We note that in the absence of a SDIPS, i.e. $\varphi
_{NN}^{\uparrow}=\varphi_{NN}^{\downarrow}$, Eq. (\ref{Thesingle}) is in
agreement with Eq. (7) of Ref.~\onlinecite{Martin-Rodero}. For $e\left|
V\right|  >\Delta$, one can calculate $G$ from Eq. (\ref{Gfns}) by using
\begin{align}
&  \mathcal{W}_{F,F}^{(e,\sigma)}(\varepsilon)\nonumber\\
&  =\frac{T_{\sigma}(1-\left|  \gamma\right|  ^{2}R_{-\sigma})(1+\left|
\gamma\right|  ^{2})}{1+\left|  \gamma\right|  ^{4}R_{\uparrow}R_{\downarrow
}-2\left|  \gamma\right|  ^{2}\sqrt{R_{\uparrow}R_{\downarrow}}\cos
[\varphi_{NN}^{\sigma}-\varphi_{NN}^{-\sigma}]} \label{Wmonocanal}%
\end{align}
for $\left|  \varepsilon\right|  >\Delta$. The expression of $S$ is too
complicated to be given here. The above Eqs. describe a phenomenon analogous
to the one predicted for a $S/F/N$ quantum point contact\cite{Zhao}. The
denominators of Eqs. (\ref{Thesingle}) and (\ref{Wmonocanal}) contain
interference terms which describe iterative reflection processes between the
$S/N$ and $N/F$ interfaces (Andreev bound states). A quasiparticle can
interfere with itself after two back and forth travels between $S$ and $F$,
one as an electron $(e,\sigma)$ and one as a hole $(h,-\sigma)$. The
conductance and noise depend on the phase difference $\Delta\varphi
=\varphi_{NN}^{\uparrow}-\varphi_{NN}^{\downarrow}$ because the $N/F$
scattering matrix for holes is $\left(  \mathcal{P}^{-\sigma}\right)  ^{\ast}%
$. This picture is in fact valid at any\ temperatures and voltages. In the
general case, the whole $\mathcal{S}$ matrix is necessary to calculate the
current and noise, but all the elements of $\mathcal{S}$ have the same
denominator as the one appearing at the right hand side of Eq.
(\ref{Thesingle}) and (\ref{Wmonocanal}) for $\left|  \varepsilon\right|
<\Delta$ and $\left|  \varepsilon\right|  >\Delta$ respectively [see Appendix
B]. In addition, we have checked analytically that, in the general case, $G$
and $S$ depend on the phases of the $N/F$ scattering matrix through the
parameter $\Delta\varphi$ only (they are actually periodic functions of
$\Delta\varphi$ with a $2\pi$-periodicity).

\begin{center}
\begin{figure}[ptb]
\includegraphics[width=1.\linewidth]{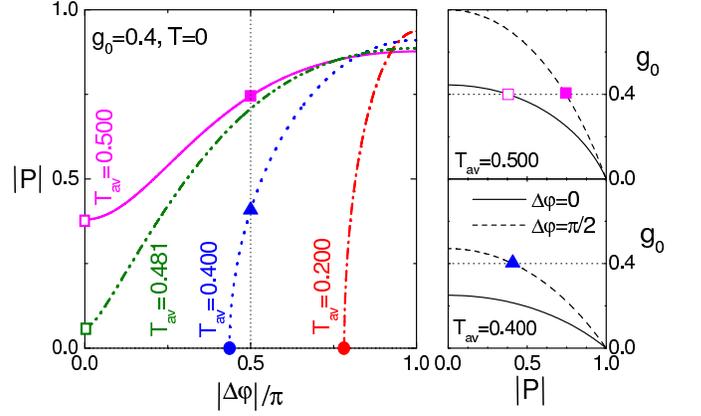}\caption{Left panel:
Polarization $\left|  P\right|  $ calculated for $g_{0}=G(V=0)h/e^{2}=0.4$ and
different values of $G(V=+\infty)=2T_{av}e^{2}/h$, as a function of the SDIPS
parameter $\left|  \Delta\varphi\right|  $, at zero temperature ($T=0$). Right
panels: Reduced zero-bias conductance $g_{0}$ of the QPC versus $\left|
P\right|  $, for $T_{av}=0.5$ (upper right panel), $T_{av}=0.4$ (bottom right
panel) and different values of $\Delta\varphi$, i.e. $\Delta\varphi=0$ (solid
lines) and $\Delta\varphi=\pi/2$ (dashed lines). In the left panel,
$T_{av}\leq0.5$ is assumed, so that the number of solutions for $\left|
P\right|  $ is either 0 or 1 depending on the value of $\left|  \Delta
\varphi\right|  $. In all the cases considered, the inferred $\left|
P\right|  $ strongly increases with $\left|  \Delta\varphi\right|  $, because
$g_{0}$ decreases with both $\left|  P\right|  $ and $-\left|  \Delta
\varphi\right|  $ for $0<\left|  \Delta\varphi\right|  <\pi$ (see right
panels). For high enough values of $T_{av}$, \ there exists a $\left|
P\right|  $ solution for $\Delta\varphi=0$ (see e.g. open pink square for
$T_{av}=0.5$ and open green square for $T_{av}=0.481$). For lower values of
$T_{av}$ , there is no $\left|  P\right|  $ solution if $\Delta\varphi=0$,
because the zero-bias conductance of the device cannot reach $g_{0}$ (see blue
dotted line for $T_{av}=0.4$ and red dot-dashed line for $T_{av}=0.2$).
However, it is possible to find a $\left|  P\right|  $ solution at $\left|
\Delta\varphi\right|  $ finite because $g_{0}$ increases with $\left|
\Delta\varphi\right|  $ for $0<\left|  \Delta\varphi\right|  <\pi$ (see e.g.
blue triangle in the bottom right panel, for $T_{av}=0.4$ and $\left|
\Delta\varphi\right|  =\pi/2$). In some cases, a finite $\Delta\varphi$ can
explain the considered values of $g_{0}$ and $T_{av}$ even in the absence of
polarization (see blue and red solid circles in the left panel).}%
\label{Pola}%
\end{figure}
\end{center}

We now focus on the zero-voltage conductance $G(V=0)=g_{0}e^{2}/h$ of the QPC,
which writes, from Eqs. (\ref{Gfns}) and (\ref{Thesingle}),%
\begin{equation}
g_{0}=\frac{4T_{\uparrow}T_{\downarrow}}{1+R_{\uparrow}R_{\downarrow}%
+2\sqrt{R_{\uparrow}R_{\downarrow}}\cos[\Delta\varphi]}\leq4 \label{g0}%
\end{equation}
In an experimental context, one may hope to determine the value of the
polarization $P=(T_{\uparrow}-T_{\downarrow})/(T_{\uparrow}+T_{\downarrow})$
from the zero voltage conductance and the high voltage conductance
$G(V=+\infty)=2T_{av}e^{2}/h$, with $T_{av}=\left(  T_{\uparrow}%
+T_{\downarrow}\right)  /2$. We define
\begin{equation}
A_{1(2)}=\left(  g_{0}\cos(\Delta\varphi)\pm D\right)  /\left(  4-g_{0}%
\right)
\end{equation}
with
\begin{equation}
D=\sqrt{16(1-2T_{av})+8g_{0}T_{av}-g_{0}^{2}\sin^{2}(\Delta\varphi)}~\text{.}%
\end{equation}
For $i\in\{1,2\}$, the polarizations $\pm P_{i}$ with
\begin{equation}
P_{i}=\frac{\sqrt{(1-T_{av})^{2}-A_{i}^{2}}}{T_{av}}%
\end{equation}
will be solutions of the problem provided $D$ is real and $0<A_{i}<1-T_{av}$.
Thus, depending on the values of $T_{av}$, $g_{0}$ and $\Delta\varphi$, there
can be either 0, 1 or 2 solutions for $\left|  P\right|  $. Importantly, the
inferred values of $\left|  P\right|  $ depend not only on the zero voltage
and normal conductances, but also strongly on $\Delta\varphi$. For simplicity,
we will consider in the following the situation $T_{av}\leq0.5$, for which one
has $A_{1}>0$ and $A_{2}<0$, so that there is either no solution for $\left|
P\right|  $ or one solution $\left|  P\right|  =\left|  P_{1}\right|  $ if
$A_{1}<1-T_{av}$. This situation is illustrated in Fig. \ref{Pola}, whose left
panel shows the calculated $\left|  P\right|  $ as a function of $\left|
\Delta\varphi\right|  $ for $g_{0}=0.4$ and different values of $T_{av}$, with
$T_{av}\leq0.5$ (the use of $\left|  \Delta\varphi\right|  $ is due to the
fact that $\left|  P\right|  $ is an even function of $\Delta\varphi$). For
the largest values of $T_{av}$ used in Fig. \ref{Pola}, left panel (see pink
solid curve and green double dot-dashed curve), a finite $\left|  P\right|  $
is found for $\Delta\varphi=0$ (see pink and green open squares). However, the
calculated $\left|  P\right|  $ can also be larger if one uses a finite
$\left|  \Delta\varphi\right|  $. This can be understood by noting that, for
the relatively low values of $T_{av}$ used in this Figure and $0<\left|
\Delta\varphi\right|  <\pi$, $g_{0}$ decreases monotonically with both
$\left|  P\right|  $ and $-\left|  \Delta\varphi\right|  $. An increase in
$\left|  \Delta\varphi\right|  $ thus compensates an increase in $\left|
P\right|  $ (see upper right panel of Fig. \ref{Pola}). For the lowest values
of $T_{av}$ used in Fig. \ref{Pola}, left panel (see blue dotted curve and red
dot-dashed curve), there is no $\left|  P\right|  $ solution if $\Delta
\varphi=0$, because the zero-bias conductance of the device cannot reach the
considered value $g_{0}=0.4$. However, as long as $g_{0}$ is not too large, it
is possible to find a $\left|  P\right|  $ solution at $\left|  \Delta
\varphi\right|  $ finite because the zero bias conductance of the QPC
increases with $\left|  \Delta\varphi\right|  $ for $0<\left|  \Delta
\varphi\right|  <\pi$ (see e.g. blue triangle in the bottom right panel of
Fig. \ref{Pola}). In this case, the calculated value of $\left|  P\right|  $
also increases with $\left|  \Delta\varphi\right|  $, for the same reason as
previously. Note that due to the continuity of the equations, there exists a
limiting case where a finite $\Delta\varphi$ can explain the considered values
of $g_{0}$ and $T_{av}$ in absence of polarization (these points are indicated
with solid circles in Fig. \ref{Pola})\cite{note2}. For $T_{av}\geq0.5$,
$g_{0}$ is not always a monotonic function of $\left|  P\right|  $, so that
there can be 2 solutions $\left|  P_{1}\right|  $ and $\left|  P_{2}\right|  $
for $\left|  P\right|  $ in some situations (not shown). We conclude that even
in the single-channel case, knowing the zero and high voltage conductances of
the QPC\ is not sufficient to determine $P$\cite{note}. We will thus consider
below the full voltage dependence of $G(V)$, which brings more information on
the system. We will also show that noise measurements prove to be very useful
to characterize unambiguously the properties of the QPC.

\begin{figure}[ptb]
\includegraphics[width=1.\linewidth]{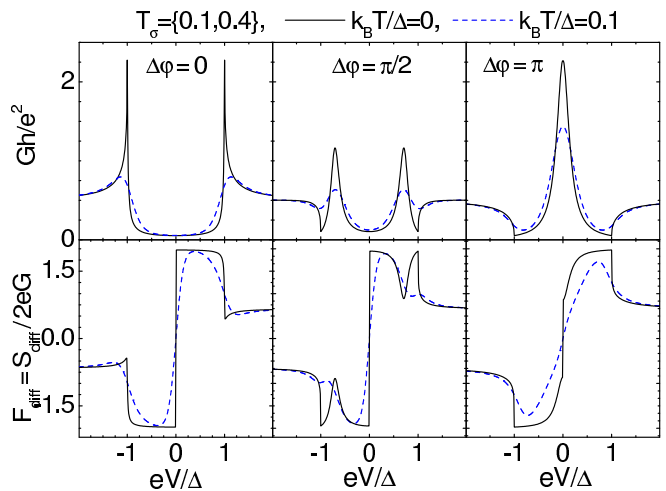}\caption{Conductance $G$
(top panels) and differential Fano factor $F_{diff}=(dS/dV)/2eG$ (bottom
panels) as a function of the bias voltage $V$ for $T_{\uparrow}%
=0.4,T_{\downarrow}=0.1$, different temperatures (black solid lines $T=0$,
blue dashed lines $k_{B}T/\Delta=0.1$), and different values of $\Delta
\varphi$ (left panels: $\Delta\varphi=0$, middle panels $\Delta\varphi=\pi/2$,
right panels $\Delta\varphi=\pi$).}%
\label{PanelLowT}%
\end{figure}

The top panels of Figs. \ref{PanelLowT} and \ref{PanelHighT} show the voltage
dependence of $G$ for low and high values of $T_{av}$, respectively. In this
paragraph, we comment on the results for $T=0$ only (black full lines). For
$\Delta\varphi=0$, the conductance shows peaks at $eV=\pm\Delta$, because for
$\varepsilon=\pm\Delta$, one has $2\varphi_{a}=0$~\textrm{[}$2\pi$\textrm{]},
so that the multiple reflection paths between $S$ and $F$ interfere
constructively. Between these peaks, the conductance reaches a minimum of
$4(e^{2}/\hbar)T_{\uparrow}T_{\downarrow}/[1+(R_{\uparrow}R_{\downarrow
})^{1/2}]^{2}$ at $V=0$. The existence of a finite SDIPS can strongly modify
this behavior. Indeed, for $\Delta\varphi\neq0$ and a small enough value of
$T_{av}$, the resonance peaks of $G(V)$ are shifted to lower voltages $V=\pm
V_{p}$ with $eV_{p}\simeq\Delta\cos(\Delta\varphi/2)<\Delta$ (Fig.
\ref{PanelLowT}, top middle panel). For $\Delta\varphi=\pi$, these peaks are
merged into a single peak and $G(V)$ is maximum at $V=0$ (Fig. \ref{PanelLowT}%
, top right panel). For high values of $T_{av}$, no clear subgap conductance
peaks occur, but the curvature of the sub-gap $G(V)$ characteristic can be
inverted by a finite SDIPS (Fig. \ref{PanelHighT}, top panels). The $G(V)$
curve is independent from $\Delta\varphi$ only if interferences are suppressed
for one spin direction, i.e. $T_{\uparrow}=1$ or $T_{\downarrow}=1$. Above the
gap, in any case, the conductance of the device drops to its normal-state
value $2\left(  e^{2}/h\right)  T_{av}$ which does not depend on the SDIPS.

The current noise through the $S/F$ QPC can provide more information on its
properties, which is of particular interest since this device is characterized
by a larger number of parameters than in the spin-degenerate case. According
to Eqs. (\ref{Gfns}) and (\ref{Sfns}), the conductance and the noise are not
mathematically equivalent and the noise can thus provide information
complementary to the conductance. The bottom panels of Figure \ref{PanelLowT}
and \ref{PanelHighT} show the voltage dependence of the differential Fano
factor $F_{diff}=(dS(V)/dV)/2eG$ for low and high values of $T_{av}$,
respectively. This quantity can be measured directly (see e.g. Ref.
\onlinecite{Schoelkopf}) or obtained from a $S(V)$ measurement. In this
paragraph, we comment on the results for $T=0$ only. One can first note that
$F_{diff}(V)$ is an odd function of $V$ due to $S(V)=S(-V)$. For a low
$T_{av}$ and $\Delta\varphi=0$, $F_{diff}$ shows subgap plateaus at values
$\pm2[(r_{\uparrow}+r_{\downarrow})/(1+r_{\uparrow}r_{\downarrow})]^{2}$ for
$V\gtrless0$, with $r_{\sigma}=(R_{\sigma})^{1/2}$. For $\Delta\varphi\neq0$,
a dip/peak appears on these plateaus, at the resonance voltages $V=\pm V_{p}$,
again due to constructive quasiparticle interferences (Fig. \ref{PanelLowT},
bottom middle panel). For higher values of $T_{av}$, the dips and peaks are
smoothed, but the shape of $F_{diff}(V)$ remains sensitive to $\varphi$ as
long as $T_{\uparrow}T_{\downarrow}<1$ (Fig. \ref{PanelHighT}). Above the gap,
in any case, $F_{diff}$ drops to the normal state value $(\sum
\nolimits_{\sigma}T_{\sigma}[1-T_{\sigma}])/(\sum\nolimits_{\sigma}T_{\sigma
})$ which does not depend on the SDIPS. One can thus determine the
polarization $P$ of the tunnel rates by using%
\[
P^{2}=\frac{2e^{2}}{h}\frac{1-F_{diff}(V=+\infty)}{G(V=+\infty)}-1
\]
Then, the SDIPS parameter $\Delta\varphi$ and the BCS\ gap $\Delta$ can be
determined from the voltage dependence of the conductance and noise curves.

\subsection{Conductance and noise of the QPC at finite temperatures}

\begin{figure}[ptb]
\includegraphics[width=1.\linewidth]{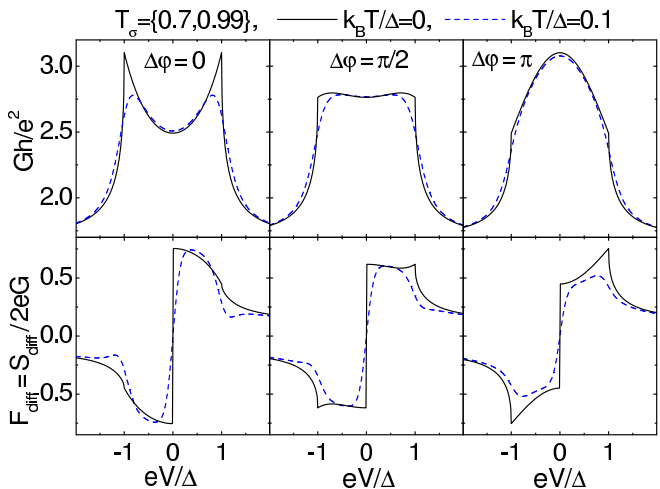}\caption{{}Conductance $G$
(top panels) and differential Fano factor $F_{diff}$ (bottom panels) as a
function of the bias voltage $V$ for $T_{\uparrow}=0.99,T_{\downarrow}=0.7$,
different temperatures (black solid lines $T=0$, blue dashed lines
$k_{B}T/\Delta=0.1$), and different values of $\Delta\varphi$ (left panels:
$\Delta\varphi=0$, middle panels $\Delta\varphi=\pi/2$, right panels
$\Delta\varphi=\pi$).}%
\label{PanelHighT}%
\end{figure}

\begin{center}
\begin{figure}[ptb]
\includegraphics[width=1.\linewidth]{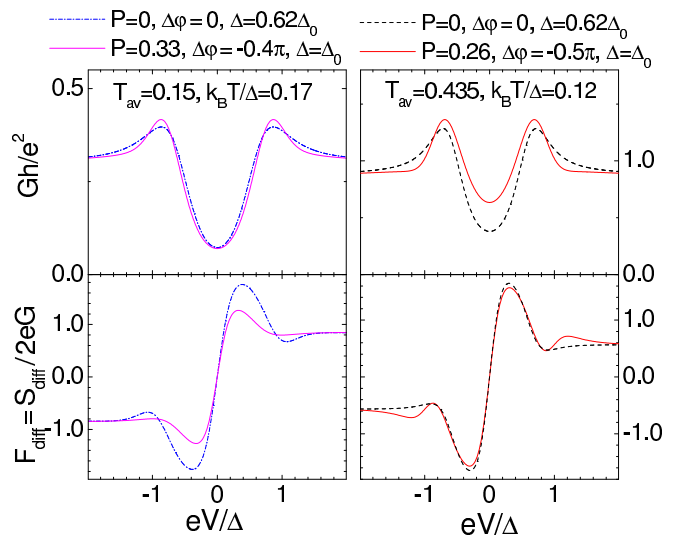}\caption{{}Conductance
$G$ (top panels) and differential Fano factor $F_{diff}$ (bottom panels) as a
function of the bias voltage $V$, for different sets of parameters. The left
panels compare a case with a finite SDIPS (pink solid lines) and a case with
no SDIPS and a smaller gap value $\Delta$ (blue dash-dotted lines). The
conductance curves obtained in the two cases (top left panels) are very close,
but the differential Fano factor curves (bottom left panels) have a strong
quantitative difference. The right panels also compare a case with a finite
SDIPS (red solid lines) and a case with no SDIPS and a smaller $\Delta$ (black
dashed lines). In these two cases, the conductance curves (top right panel)
are qualitatively similar but the Fano factor curves (bottom right panel) show
a strong qualitative difference: in the case of a finite SDIPS, $F_{diff}(V)$
shows a secondary peak[dip] at $eV>\Delta$ [$eV<\Delta$]. This feature never
appears in the absence of a SDIPS. Note that the high voltage limits of the
two $F_{diff}(V)$ curves shown in the bottom left (bottom right) panel are
different, but that this difference is not visible on the scale shown in the
figure.}%
\label{Comparaisons}%
\end{figure}
\end{center}

We now comment on the finite temperature $G(V)$ and $F_{diff}(V)$ curves,
obtained by differentiating Eqs. (\ref{Igal}) and (\ref{Sgal}) with respect to
$V$ (see blue dashed lines in Figs. \ref{PanelLowT} and \ref{PanelHighT}). At
first glance, these curves simply seem to be thermally rounded. However, the
finite temperature expressions of $G(V)$ and $F_{diff}(V)$ involve elements of
the $\mathcal{S}$ matrix which do not appear in the zero-temperature
expressions, thus the finite temperature curves sometimes display features not
predictable from the zero temperature curves. For instance, in the bottom left
panel of Fig. \ref{PanelHighT}, a slight dip[peak] appears in the
$F_{diff}(V)$ curve at $V=[-]\Delta/e$ for $T\neq0$, an effect not present for
$T=0$. In the case of a weakly transparent contact, since the Andreev
resonance peaks of the conductance are shifted to lower voltages for
$\Delta\varphi\neq0$, one can obtain, for $T\neq0$, conductance curves similar
to those obtained for $\Delta\varphi=0$ and a reduced value of $\Delta$. The
determination of the quantum point contact properties from the $G(V)$ curves
alone can thus be difficult at finite temperatures. Figure \ref{Comparaisons}
shows two examples where measuring the voltage dependence of the noise can
clearly bring more information on the system. The left panel presents two
cases where the $G(V)$ curves are extremely close, one case with
$\Delta\varphi\neq0$ and one case with $\Delta\varphi=0$ and a reduced gap
value. The corresponding $F_{diff}(V)$ curves have a strong quantitative
difference, which can help to discriminate the two cases. The right panels of
Fig. \ref{Comparaisons} present two cases where the $G(V)$ curves are
qualitatively similar, one case with $\Delta\varphi\neq0$ and one case with
$\Delta\varphi=0$ and a reduced gap value. The corresponding $F_{diff}(V)$
curves have a strong \textit{qualitative} difference, which can help to
discriminate the two cases again. In the case $\Delta\varphi\neq0$, the
differential Fano factor shows a secondary peak[dip] at $eV>0$ [ $eV<0$]. This
effect, which can also be seen in the bottom middle panel of Fig.
\ref{PanelLowT}, never occurs for $\Delta\varphi=0$.

Before concluding this section, we point out that in most cases, experimental
results on $S/F$ QPCs were interpreted with various models inspired from the
BTK approach\cite{Soulen}. The effect of the SDIPS was not studied in these
works. However the description of a $S/F$ interface with a delta-function
barrier naturally takes into account the SDIPS. We have checked that the
spin-dependent BTK-like model corresponds to a particular case of the
scattering model of Sec. II (generalized to the multichannel case), with
parameters given in appendix C.

\section{Conclusion and discussion}

In this paper, we have studied the conductance $G$ and noise $S$ of a
superconducting/ferromagnetic ($S/F$) single channel Quantum Point Contact
(QPC) as a function of the QPC bias voltage $V$, using a scattering approach.
We have shown that the Spin-Dependence of Interfacial Phase Shifts (SDIPS)
acquired by electrons at the $S/F$ interface strongly modifies these signals.
In particular, for a weakly transparent contact, the SDIPS produces unusual
sub-gap resonances or dips in the $G(V)$ and $S(V)$ curves. We have shown that
measuring the noise should help to gain information on the system. This fact
is well illustrated e.g. by cases where the SDIPS modifies qualitatively the
zero-frequency noise but not the conductance.

One should note that, so far, experiments on $S/F$ quantum point contacts were
performed in the multichannel limit. Although single-channel contacts were
already realized in the $S/N$ case\cite{Scheer}, this seems more difficult in
the $S/F$ case with the present fabrication techniques. A multichannel theory
should thus be very useful. For a multichannel disordered $S/N$ QPC, Ref.
\onlinecite{BeenakkerSNN} has shown that, due to time reversal symmetry, the
eigenstates the QPC normal state scattering matrix are also those of
$\mathcal{S}$. As a consequence, the multichannel generalization of $G(V)$ and
$S(V)$ just requires a summation on the normal channels index. In the $S/F$
case, the calculation should be more complicated because time reversal
symmetry breaking suppresses the normal channels independence
\cite{Martin-Rodero,Xia}. So far, the multichannel case has been addressed
with the BTK model, in which the hypothesis of a specular interface allows one
to consider the normal channels as independent in spite of the time reversal
symmetry breaking\cite{DeJong,Zutic}. However, this approach lacks of
generality since it imposes a particular relation between the transmission
probabilities and the SDIPS parameters, and it is not valid for disordered
interfaces. Interestingly, Refs. \onlinecite{Xia} reports on the ab-initio
modelization of a particular type of disordered multichannel $S/F$ interface.
However, to our knowledge, the multichannel $S/F$ case has not been studied
with a more general approach so far. Our work suggests that such a study
should take into account the SDIPS. Considering that, even in the single
channel case, the conductance and noise of the device vary nonlinearly with
the SDIPS parameter (see e.g. Eq. \ref{Thesingle}), there is no reason to
expect that the effects of the SDIPS average out in the multichannel case,
even in the extreme case of a SDIPS parameter $\Delta\varphi^{n}$ randomly
distributed with the channel index $n$.

We would like to acknowledge useful discussions with C. Bruder, B. Dou\c{c}ot
and T. Kontos. This work was financially supported by the Swiss National
Science Foundation, the DFG through SFB 513 and SFB 767, the Priority Programm
\textit{Semiconductor Spintronics}, and the Landesstiftung Baden-W\"{u}%
rttemberg through the Kompetenznetzwerk \textit{Funktionelle Nanostrukturen}.

\section{Appendix A : A symmetry property of the scattering matrix
$\mathcal{S}$}

This appendix shows that the property $\mathcal{T}_{M_{1},M_{2}}^{\alpha
,\beta}(-\varepsilon)=\mathcal{T}_{M_{1},M_{2}}^{\widetilde{\alpha}%
,\widetilde{\beta}}(\varepsilon)$, with $(\alpha,\beta)\in\mathcal{E}_{\sigma
}^{2}$ and $(M_{1},M_{2})\in\{S,F\}^{2}$, used to derive Eqs. (\ref{Gfns}) and
(\ref{Sfns}) is a general property which stems from the symmetries of the BdG equations.

We first consider the eigenstates of the BdG equations in a bulk S. An
eigenstate with energy $\varepsilon$ in the subspace $\mathcal{E}_{\sigma
}=\{(e,\sigma),(h,-\sigma)\}$ has electron and hole components $u_{\sigma}$
and $v_{-\sigma}$ such that
\[
\left[
\begin{array}
[c]{cc}%
H_{\sigma} & \Delta\\
\Delta^{\ast} & -H_{-\sigma}^{\ast}%
\end{array}
\right]  \left[
\begin{array}
[c]{c}%
u_{\sigma}\\
v_{-\sigma}%
\end{array}
\right]  =\varepsilon\left[
\begin{array}
[c]{c}%
u_{\sigma}\\
v_{-\sigma}%
\end{array}
\right]
\]
with $H_{\sigma}$ the normal state hamiltonian for electrons with spin
$\sigma$. This equation can be rewritten as
\[
\left[
\begin{array}
[c]{cc}%
H_{-\sigma} & \Delta\\
\Delta^{\ast} & -H_{\sigma}^{\ast}%
\end{array}
\right]  \left[
\begin{array}
[c]{c}%
-v_{-\sigma}^{\ast}\\
u_{\sigma}^{\ast}%
\end{array}
\right]  =-\varepsilon\left[
\begin{array}
[c]{c}%
-v_{-\sigma}^{\ast}\\
u_{\sigma}^{\ast}%
\end{array}
\right]  ~\text{.}%
\]
This shows that $^{t}[-v_{-\sigma}^{\ast},~u_{\sigma}^{\ast}]$ is another
solution of the BdG Equations, with energy $-\varepsilon$ in the
$\mathcal{E}_{-\sigma}$ subspace. This property is also valid in a bulk $F$
(in this case $\Delta=0$ and $v_{-\sigma}=0$).

We now consider the scattering processes at a $S/F$ interface. From the
definition of $\mathcal{S}$, and outgoing wave $\Psi_{M_{1},\sigma}^{o}%
=^{t}(a_{M_{1},\sigma}^{o}~b_{M_{1},\sigma}^{o})$ with energy $\varepsilon$ in
lead $M\in\{S,F\}$ writes%
\[
\Psi_{M_{1},\sigma}^{o}=\sum\limits_{M_{2}}\left[
\begin{array}
[c]{cc}%
\mathcal{S}_{M_{1},M_{2}}^{(e,\sigma),(e,\sigma)}(\varepsilon) &
\mathcal{S}_{M_{1},M_{2}}^{(e,\sigma),(h,-\sigma)}(\varepsilon)\\
\mathcal{S}_{M_{1},M_{2}}^{(h,-\sigma),(e,\sigma)}(\varepsilon) &
\mathcal{S}_{M_{1},M_{2}}^{(h,-\sigma),(h,-\sigma)}(\varepsilon)
\end{array}
\right]  \Psi_{M_{2},\sigma}^{i}%
\]
with $\Psi_{M_{2},\sigma}^{i}=^{t}[a_{M_{2},\sigma\text{,}}^{i}~b_{M_{2}%
,\sigma}^{i}]$ the incoming state with energy $\varepsilon$ in the
$\mathcal{E}_{\sigma}$ space of lead $M_{2}$. This induces%
\begin{align}
&  \Phi_{M_{1},\sigma}^{o}\nonumber\\
&  =\sum\limits_{M_{2}}\left[
\begin{array}
[c]{cc}%
\left(  \mathcal{S}_{M_{1},M_{2}}^{(h,-\sigma),(h,-\sigma)}(\varepsilon
)\right)  ^{\ast} & -\left(  \mathcal{S}_{M_{1},M_{2}}^{(h,-\sigma
),(e,\sigma)}(\varepsilon)\right)  ^{\ast}\\
-\left(  \mathcal{S}_{M_{1},M_{2}}^{(e,\sigma),(h,-\sigma)}(\varepsilon
)\right)  ^{\ast} & \left(  \mathcal{S}_{M_{1},M_{2}}^{(e,\sigma),(e,\sigma
)}(\varepsilon)\right)  ^{\ast}%
\end{array}
\right]  \Phi_{M_{2},\sigma}^{o} \label{R}%
\end{align}
with $\Phi_{M_{k},\sigma}^{o[i]}=^{t}[-(b_{M_{k},\sigma}^{o[i]})^{\ast
},~(a_{M_{k},\sigma}^{o[i]})^{\ast}]$ for $k\in\{1,2\}$. From the previous
paragraph, $\Phi_{M_{k},\sigma}^{o[i]}$ corresponds to an outgoing [incoming]
state with energy $-\varepsilon$ in the $\mathcal{E}_{-\sigma}$ subspace of
lead $M_{k}$. Equation (\ref{R}) thus gives $\mathcal{S}_{M_{1},M_{2}%
}^{(e,-\sigma),(e,-\sigma)}(-\varepsilon)=\left(  \mathcal{S}_{M_{1},M_{2}%
}^{(h,-\sigma),(h,-\sigma)}(\varepsilon)\right)  ^{\ast}$ and $\mathcal{S}%
_{M_{1},M_{2}}^{(h,\sigma),(e,-\sigma)}(-\varepsilon)=-\left(  \mathcal{S}%
_{M_{1},M_{2}}^{(e,\sigma),(h,-\sigma)}(\varepsilon)\right)  ^{\ast}$, which
generalizes Eq. (C7) of Ref.~\onlinecite{Datta} to the spin-dependent case.
The relation $\mathcal{T}_{M_{1},M_{2}}^{\alpha,\beta}(-\varepsilon
)=\mathcal{T}_{M_{1},M_{2}}^{\widetilde{\alpha},\widetilde{\beta}}%
(\varepsilon)$ follows straightforwardly.

\section{Appendix B : Coefficients of the $F/S$ scattering matrix}

When a $S/F$ interface can be modeled as a ballistic $S/N$ interface in series
with a dirty $N/F$ interface, with the thickness of $N$ tending to zero, the
scattering matrix $\mathcal{S}(\varepsilon)$ of the $S/F$ interface can
expressed in terms of the scattering matrix $\mathcal{P}^{\sigma}%
(\varepsilon)$ of electrons with spins $\sigma$ on the $N/F$ interface and of
the Andreev reflection amplitude $\gamma$ defined in Section II.B. One finds
Eqs. (\ref{S2}-\ref{S4}), and
\begin{equation}
\mathcal{S}_{S,S}^{(h,-\sigma),(e,\sigma)}(\varepsilon)=\gamma\lbrack
\gamma_{t}^{2}\left(  \mathcal{P}_{NN}^{-\sigma}(-\varepsilon)\right)  ^{\ast
}M^{\sigma}\mathcal{P}_{NN}^{\sigma}(\varepsilon)-\mathbb{I}_{S,(e,\sigma
)}]\nonumber
\end{equation}%
\begin{align*}
\mathcal{S}_{S,F}^{(e,\sigma),(e,\sigma)}(\varepsilon)  &  =\gamma
_{t}\mathcal{P}_{NF}^{\sigma}(\varepsilon)\\
&  +\gamma_{t}\gamma^{2}\mathcal{P}_{NN}^{\sigma}(\varepsilon)N^{\sigma
}\left(  \mathcal{P}_{NN}^{-\sigma}(-\varepsilon)\right)  ^{\ast}%
\mathcal{P}_{NF}^{\sigma}(\varepsilon)
\end{align*}%
\begin{align*}
\mathcal{S}_{F,S}^{(e,\sigma),(e,\sigma)}(\varepsilon)  &  =\gamma
_{t}\mathcal{P}_{FN}^{\sigma}(\varepsilon)\\
&  +\gamma^{2}\gamma_{t}\mathcal{P}_{FN}^{\sigma}(\varepsilon)N^{\sigma
}\left(  \mathcal{P}_{NN}^{-\sigma}(-\varepsilon)\right)  ^{\ast}%
\mathcal{P}_{NN}^{\sigma}(\varepsilon)
\end{align*}%
\[
\mathcal{S}_{F,S}^{(h,-\sigma),(e,\sigma)}(\varepsilon)=\gamma\gamma
_{t}\left(  \mathcal{P}_{FN}^{-\sigma}(-\varepsilon)\right)  ^{\ast}M^{\sigma
}\mathcal{P}_{NN}^{\sigma}(\varepsilon)
\]%
\begin{equation}
\mathcal{S}_{S,F}^{(h,-\sigma),(e,\sigma)}(\varepsilon)=\gamma\gamma
_{t}\left(  \mathcal{P}_{NN}^{-\sigma}(-\varepsilon)\right)  ^{\ast}M^{\sigma
}\mathcal{P}_{NF}^{\sigma}(\varepsilon)\nonumber
\end{equation}%
\[
\mathcal{S}_{S,S}^{(e,\sigma),(e,\sigma)}(\varepsilon)=\gamma_{t}%
^{2}\mathcal{P}_{NN}^{\sigma}(\varepsilon)N^{\sigma}%
\]
with $\gamma_{t}=[1-\left|  \gamma\right|  ^{2}]^{1/2}$ the Andreev
transmission amplitude. The eight missing elements of $\mathcal{S}%
(\varepsilon)$ can be obtained from the above Eqs. by replacing $(h,-\sigma)$
by $(e,\sigma)$ and vice versa in the upper indices of the $\mathcal{S}$
elements and by doing the permutations $\mathcal{P}^{\sigma}(\varepsilon
)\leftrightarrows\mathcal{P}^{-\sigma}(-\varepsilon)^{\ast}$ and $N^{\sigma
}\leftrightarrows M^{\sigma}$ at the right hand sides of the equations.

\section{Appendix C : Equivalent parameters of the BTK model}

In some works about QPCs \cite{Soulen}, the data were interpreted in terms of
a generalization of the BTK model to the $S/F$ case. In this approach, no
fictitious $N$ layer is used. The $S$ and $F$ leads are described with BdG
equations\cite{BdG}, and the $S/F$ interfacial scattering is attributed to a
delta-function barrier $V_{\sigma}(x)=H_{\sigma}\delta(x)$. A multichannel
description is generally used, where channel $n$ corresponds to the $n^{th}$
transverse mode of the device, for which quasiparticles have a spin-dependent
wavevector $\pm k_{\sigma}^{n}$ in $F$ and a spin-independent wavevector $\pm
q^{n}$ in $S$ with $k_{\sigma}^{n}=[(2m/\hbar^{2})(E_{F}^{F}-E_{n}+\sigma
E_{ex})]^{1/2}$, $q^{n}=[(2m/\hbar^{2})(E_{F}^{S}-E_{n})]^{1/2}$,
$E_{F}^{S(F)}$ the Fermi level in $S(F)$, $E_{ex}$ the exchange field in $F$
and $E_{n}$ the energy of the $n^{th}$ transverse mode\cite{DeJong}. Due to
the spectular nature of the $S/F$ interface, the scattering matrix
$\mathcal{S}(\varepsilon)$ associated to the BTK-like model does not connect
the different transverse modes of the device. Consequently, the conductance
and noise of the QPC can be obtained by summing the expressions introduced in
this article on the channel index $n$. The scattering parameters associated to
channel $n$ are the transmission probability
\[
T_{\sigma}^{n}=4k_{\sigma}^{n}q^{n}/[\left(  k_{\sigma}^{n}+q^{n}\right)
^{2}+K_{\sigma}^{2}]
\]
and the SDIPS parameter
\[
\Delta\varphi^{n}=\arg[b_{\uparrow}^{n}/b_{\downarrow}^{n}]
\]
with $K_{\sigma}=2mH_{\sigma}/\hbar^{2}$, $m$ the effective mass of the
particles, and
\[
b_{\sigma}^{n}=(q^{n}-k_{\sigma}^{n}-\mathbf{i}K_{\sigma})/(k_{\sigma}%
^{n}+q^{n}+\mathbf{i}K_{\sigma})
\]
Interestingly, from the above Eqs., one can notice that, in principle,
$\Delta\varphi^{n}$ can be finite even if $K_{\sigma}$ is not spin-dependent,
provided $k_{\sigma}^{n}$ is spin-dependent and $K_{\sigma}$ is not too large.


\begin{thebibliography}{99}
\bibitem{NoiseSF}We are only aware of a brief zero-voltage BTK-like
calculation of the noise through a $S/F$ QPC\cite{DeJong} and a
two-dimensional tight-bindind calculation of the current cross-correlations
between two $F$ leads connected through a $S$ electrode\cite{Taddei2}.
\bibitem{comp}Equation (\ref{Sfns}) is a generalization of Eq. (54) of Ref.
\onlinecite{Anantram} to the spin-dependent case and both signs of V.
\bibitem{Vdrop}In the model introduced in section II.B, the voltage drop due
to $V$ can be considered as located on the $F/N$ interface, so that the
scattering matrix of the fictitious $N/S$ interface does not depend on $V$.
Then, if $R_{\sigma}$, $T_{\sigma}$ and $\varphi_{ij}^{\sigma}$ are assumed to
be independent from $\varepsilon$ and $V$, $\mathcal{W}_{F,F}^{\alpha}$, does
not depend on $V$.
\bibitem{note2}The situation $P=0$ and $\Delta\varphi\neq0$ can happen e.g.
for specific parameters in a delta function barrier model, see Fig.1 of Ref.
\onlinecite{CottetEurophys}.
\bibitem{note}We will not discuss on how to relate the polarization $P_{bulk}$
of the density of states of $F$ to the polarisation $P$ of the transmissions
discussed by us. It has been shown that this is not straightforward
\cite{Xia,Taddei}.
\end{thebibliography}
\end{document}